\documentclass[12pt]{article}
\usepackage{amsmath}
\usepackage{amsfonts}

\newtheorem{theorem}{Theorem}

\newtheorem{example}[theorem]{Example}
\newtheorem{lemma}[theorem]{Lemma}

\newcommand{\R}{{\mathbf R}}
\newcommand{\gr}{} 
\newcommand{\re}{} 

\newcommand{\p}{\partial}

\newcommand{\ds}{\displaystyle}

\newcommand{\Hess}{\mathop{\rm Hess}}
\newcommand{\Rc}{{\rm Ric}}
\newcommand{\Sec}{{\rm Sec}}
\newcommand{\vg}{{{\rm vol}_g}}

\newcommand{\Tr}{\mathop{\rm Tr}}

\newcommand{\CD}{\mathop{\rm CD}}
\newcommand{\TCD}{\mathop{\rm TCD}}

\begin{document}

\author{
Robert McCann\thanks{
Department of Mathematics,
University of Toronto, Toronto Ontario M5S 2E4 Canada,
{\tt mccann@math.toronto.edu}}}

\title{Trading linearity for ellipticity: a nonsmooth approach to Einstein's theory of gravity and the Lorentzian splitting theorems\thanks{
RJM acknowledges partial support of his research by the Canada Research Chairs program CRC-2020-00289, a grant from the Simons
Foundation (923125, McCann) and Natural Sciences and Engineering Research Council of Canada Grant RGPIN-2020-04162. 
\copyright 2024 by the author.}}
\date{\today}

\maketitle

\begin{abstract}
While Einstein's theory of gravity is formulated in a smooth setting, the celebrated singularity theorems of Hawking and Penrose describe many physical situations in which this smoothness must eventually break down. In positive-definite signature, there is a highly successful theory of metric and metric-measure geometry which includes Riemannian manifolds as a special case, but permits the extraction of nonsmooth limits under dimension and curvature bounds analogous to the energy conditions from relativity: here sectional curvature is reformulated through triangle comparison, while Ricci curvature is reformulated using entropic convexity along geodesics of probability measures.

This lecture highlights recent progress in the development of an analogous theory in Lorentzian signature, whose ultimate goal is to provide a nonsmooth theory of gravity. In particular, we foreshadow a low-regularity splitting theorem obtained by sacrificing linearity of the d'Alembertian to recover ellipticity. We exploit a negative homogeneity 
$p$-d'Alembert operator for this purpose. The same technique yields a simplified proof of Eschenberg (1988), Galloway (1989), and Newman's (1990) confirmation of Yau's (1982) conjecture, bringing 
both Lorentzian splitting results into a framework closer to the Cheeger--Gromoll (1971) splitting theorem from Riemannian geometry.
 \end{abstract}

\bigskip


\subsection*{Introduction}

In 2022, the Fields Institute sponsored a thematic semester on Nonsmooth Riemannian and Lorentzian geometry which I co-organized.
That semester featured a graduate course by McMaster Dean's Distinguished Visiting Professor Nicola Gigli (SISSA) on his nonsmooth Riemannian splitting theorem~\cite{Gigli13+},  
and --- thanks to supplemental funding
available postpandemic --- ten postdoctoral fellows including Dr.\ Mathias Braun,  who stayed at University of Toronto from 2022--24 before accepting a position at Switzerland's  EPFL. It also attracted numerous long and short-term visitors,  including former Toronto postdoctoral fellow Clemens S\"amann and 
four graduate students who have since defended their doctorates in Vienna --- 
Tobias Beran, Matteo Calisti, Argam Ohanyan and Felix Rott --- three of whom had at the time proved a nonsmooth
Lorentzian splitting theorem with Didier Solis \cite{BeranOhanyanRottSolis23} under timelike sectional curvature bounds.
This lecture is devoted to results obtained by two large research teams which coalesced during that semester:
an octet   \cite{BeranOctet24+} which developed a first-order calculus and notion of infinitesimal Minkowskianity for nonsmooth theories of gravity ---  
as well as a comparison theorem for negative homogeneity $p$-d'Alembert operators
which was novel even in the smooth context ---  and a quintet  \cite{BraunGigliMcCannOhanyanSaemann24+}
which used this idea to give a simple, new, self-contained approach to the Lorentzian splitting theorems under timelike Ricci curvature bounds.   
This research seems especially appropriate to report in the
{\em Forward from the Fields Medal 2024 Proceedings} not only because of its genesis at Toronto's Fields Institute,  but also because of the number of former Fields'
Medallists whose work impinges on this topic.

We begin with an example that illustrates what a splitting theorem is --- essentially a dimension reduction technique.

\begin{example}[When do convex functions split?]
If the graph of a {\re convex} function $u:{\gr \R^n}\longrightarrow \R$ contains a full line,
say $u (t,0,\ldots,0)=0$ for all $t \in \R$,  then $u(x) = U(x_2,\ldots,x_n)$ for all $x = (x_1,\ldots,x_n) \in \R^n$.
\end{example}

Note that the previous example requires no smoothness hypotheses.
A more sophisticated example is the celebrated splitting theorem of Cheeger and Gromoll \cite{CheegerGromoll71},
which generalized earlier results of Cohn-Vossen (for $n=2$) \cite{Cohn-Vossen36} and Toponogov (for $n \ge 2$) \cite{Toponogov64}
by substituting Ricci nonnegativity for sectional curvature nonnegativity:

\begin{example}[When do smooth Riemannian manifolds split? \cite{CheegerGromoll71}]
If a connected {\gr complete} {\re Ricci nonnegative} Riemannian manifold $(M^n,g_{ij})$ 
contains an isometric copy of a line $(\R,dr^2)$,  then $M$ is a geometric product of $\re (\R,dr^2)$ 
with a Ricci nonnegative submanifold $\gr (\Sigma^{n-1},h_{ij} = g_{ij}|_{\Sigma})$:
i.e.  there is an isometry  $(r,y) \in \R \times \Sigma \mapsto x(r,y) \in M$ with $g_{ij} dx^i dx^j = {\re dr^2} + \gr h_{kl} dy^k dy^l$.
\end{example}

Much more recently,  a nonsmooth version of this theorem has been proved in {\gr infinitesimally Hilbertian} metric-measure spaces $(M,d,{\re m})$ \cite{Gigli15}
by Gigli~\cite{Gigli13+}, assuming they satisfy a curvature-dimension condition $\CD(0,N)$ defined by Sturm \cite{Sturm06b}, Lott and Villani \cite{LottVillani09}
using a notion of entropic displacement convexity inspired by \cite{McCann97}. 
Although our primary goal is to discuss the Lorentzian analogs of such splitting theorems relevant to Einstein's theory of gravity \cite{BeemEhrlichEasley96},
let us first sketch a proof of the Cheeger--Gromoll theorem to illustrate the ideas upon which it is based.

Proof sketch:  Let $\gamma:\R\longrightarrow M^n$ be the isometrically embedded line.  Following \cite{Busemann32},
we define the Busemann functions $\pm b^\pm :=\displaystyle \lim_{r \to \pm \infty} b_r$ as limits of
$$
b_r(x):=d(x,\gamma(r)) {\gr -d(\gamma(0),\gamma(r))};
$$ 
here $b_r$ is $1$-Lipschitz and $|\nabla b_r|=1=|\nabla b^\pm|$ a.e.  For $r>0$,
the triangle inequality gives 
\begin{equation}\label{order1}
b_r \ge b^+ \ge b^- \ge -b_{-r},
\end{equation}
with all four functions vanishing at $x=\gamma(0)$.  Our second ingredient is Calabi's 
 {\em Laplacian comparison theorem} \cite{Calabi58v}, which asserts that $\Rc \ge 0$ implies
\begin{equation}\label{Laplacian comparison}
\Delta b_r = \nabla \cdot (\nabla b_r)\le \frac{n-1}{d(\cdot, \gamma(r))}
\end{equation} 
holds not only where $b_r$ is smooth, but also across the cut-locus in the {\em support sense} introduced by Calabi,
more familiar in certain communities as the {\em viscosity sense} \cite{CrandallIshiiLions92}.  Taking the limit $r\to \infty$ in
\eqref{Laplacian comparison} yields both $\pm b^\pm$ superharmonic: $\Delta b^+ \le 0 \le \Delta b^-$.  Since 
\eqref{order1} holds with equality at $x=\gamma(0)$,
the strong maximum principle gives $b^+=b^-$ hence both functions are harmonic and smooth throughout $M$.  Now Bochner's identity \cite{Bochner46} 
\begin{equation}\label{Bochner}
\gr \Tr[(\Hess b)^2] + \Rc(\nabla b,\nabla b) =  \Delta \frac{\re |\nabla b|^2}2 - g(\nabla b,\nabla {\re\Delta b})  =0
\end{equation}
yields $\Hess b =0$ for $b:=b^\pm$ since $\Rc \ge 0$.  This shows in particular that $\nabla b$ is a Killing vector field (its flow gives a local isometry)
and $\Sigma := \{ x\in M^n \mid b(x)=0\}$ is totally geodesic since its unit  normal $\nabla b$ is parallel.  Along $\Sigma$, the metric thus splits 
into tangent $g_{ij} dy^i dy^j$ and normal components $\re dr^2$.  The local isometry $(r,y) \in \R \times \Sigma \mapsto \exp_y r\nabla b(y)$ is surjective,
hence gives the global isometry desired.
\hfill $\square$

\subsection*{General relativity: Einstein's gravity and field equation}

Because space and time are intertwined in Einstein's theory of special relativity,
his theory of gravity --- general relativity --- is formulated on a smooth Lorentzian manifold. 
However,  it often predicts such manifolds are geodesically incomplete or cannot remain smooth ---
due to phenomena like black holes and the big bang.  This is a feature 
rather than a bug.

The premise of the theory --- encapsulated in the Einstein field equation \eqref{EFE} below ---
is that gravity is not a force, but rather a manifestation of curvature in the underlying geometry of spacetime.
Wheeler \cite{Wheeler98} summarized this equation with the phrase 
``Matter tells spacetime how to curve; spacetime tells matter how to move.''   In symbols,  Einstein replaced Newton's
equation relating the mass density 
\begin{equation}\label{Newton's law}
\Delta \phi = \rho \ge 0
\end{equation}
to the gravitational force $F=-\nabla\phi$ by
\begin{eqnarray}
 \re geometry &=& \re physics
\nonumber \\  \mbox{curvature} &=& \mbox{flux of energy and momentum} 
\nonumber \\ \re \Rc_{ij} - \frac12 R g_{ij} &=& \re 8\pi T_{ij} 
\label{EFE}
\end{eqnarray}
relating the geometry encoded in the signature $(+,-,-,-)$ metric tensor $g_{ij}$ to the stress-energy tensor $(T_{ij})_{i,j=0}^3$,
which measures the flux 
of $x^i$-momentum 
 in the $x^j$-direction
(substituting energy for momentum when $i=0$ and density for flux when $j=0$).
Here $\Rc_{ij}$ denotes the Ricci curvature tensor of the Lorentzian metric $g_{ij}$, whose trace   $R=g^{ij}\Rc_{ij}$ 
is the scalar curvature, and which itself is the trace $\Rc_{ij} = g^{kl}R_{ikjl}$ of the Riemann tensor described below.  As usual, summation on repeated indices is intended.

The consequences of the Einstein field equation are also illustrated by a thought experiment described by Kip Thorne \cite{Thorne94}.
Imagine you are the pilot of a spaceship sent to investigate the spacetime geometry of a spherically symmetric black hole.  You place your ship into a circular orbit
with your feet pointing toward the black hole and your head away from it.  As you gently fire your thrusters to lower the level of the orbit then --- long
before you are anywhere near the horizon of the black hole if its mass is sufficiently large  --- you begin to feel stretched from head-to-toe,  and compressed
from side-to-side and back-to-front.   This is because your head and feet are both trying to follow straight timelike geodesics into the future,  while the curvature of spacetime
due to the mass of the black hole causes these initially parallel geodesics to separate.  
Assuming you and your spaceship are very light,  so the stress-energy tensor essentially vanishes locally, the time-time component
$$
 {R_{010}}^1 + {R_{020}}^2 + {R_{030}}^3=\Rc_{00} = 0
$$ 
 of the Einstein equation asserts the front-to-back and side-to-side compression --- which have the same sign and magnitude ${R_{010}}^1= {R_{020}}^2$
by symmetry --- must be opposite in sign and half as strong as the head-to-toe stretching. 

In Newton's theory, one solves \eqref{Newton's law} to deduce the gravitational force given the mass density $\rho(x)$. 
Similarly,  if one knew the stress-energy tensor $T_{ij}(x)$ globally one might in principle solve the nonlinear system \eqref{EFE} to find the geometry $g_{ij}$.
More typically,  one knows only the stress-energy tensor in the past,  or perhaps on a spacelike slice of spacetime called a Cauchy surface.
Knowing the second fundamental form which encodes how this surface bends as it is embedded in the ambient space,  
one can then try to find the evolution of the system by solving the initial value problem as in, e.g.
Choquet-Bruhat \cite{Choquet-Bruhat52} with Geroch \cite{Choquet-BruhatGeroch69}.   This is a nonlinear wave equation, whose linearization produces gravity waves;
like other wave equations,  it is expected to propagate singularities rather than to smooth them.   On the other hand,  if one does not know the stress-energy
tensor $T_{ij}$ reflecting the physical content of the system, one can instead try to make predictions about all possible spacetime geometries which are consistent
with one or another of the conditions encoding the expected positive-definiteness properties of $T_{ij}$ locally.  Before reviewing these energy conditions,
let us recall further aspects of Lorentzian geometry.

\subsection*{Special relativity: elliptic vs hyperbolic geometry}

Euclid's geometry is based on equipping $v \in \R^n$ with the usual elliptic norm $\re |v|_E := (\sum v_i^2)^{1/2}$;  it satisfies the triangle inequality
\begin{equation}\label{norm inequality}
|v +w|_E \le |v|_E + |w|_E.
\end{equation}
The blend of space and time required by Einstein's theory of special relativity is instead set in Minkowski space, which amounts to equipping  $\R^n$ with the 
analogous {\em `hyperbolic norm'} 
\begin{equation}\label{hyperbolic norm}
 |v|_F := \begin{cases}
 (v_1^2 - \ds \sum_{i \ge 2} v_i^2)^{1/2} 
 & v \in \gr F:=\left\{v \in \R^n \mid v_1 \ge \ds 
 (\sum_{i\ge 2} v_i^2)^{1/2}
 \right\} \\
-\infty & else;
\end{cases}
\end{equation}
being concave and $1$-homogeneous,  it satisfies the backward triangle inequality
$$
|v + w |_F \ge |v|_F + |w|_F
$$ 
for all $v,w \in \R^n$,
but is terribly asymmetric: $\|-v\| \ne \|v\|$ unless $v=0$.
This asymmetry reflects our everyday experience that time always flows forward, never backward.
The convex cone $F \subset \R^n$ defined by \eqref{hyperbolic norm} is called the future cone;  a vector
$v  \in F$ is called  {\em \gr causal} or {\em \gr future-directed};
it is called {\em timelike} if $v \in F \setminus \p F$;
{\em lightlike (or null)} if $v \in \p F \setminus \{0\}$;
({\em spacelike} iff $\pm v \not\in F$ and  {\em past-directed} if $-v \in F$, though the latter two notions are not needed here).
Smooth {\gr curves} are called {\em timelike (etc.)} if all tangents are {timelike (etc.)}.

\subsection*{A crash course in differential geometry}

Consider a connected manifold $M^n$ 
and symmetric nondegenerate $C^k$-smooth tensor field $g_{ij}=g_{ji}$. 
The manifold is called {\em Riemannian} if $g_{ij}$ is positive definite at each point,  hence defines 
a Euclidean norm $| \cdot |_{E_g}$ on each tangent space.  In this case, the geometry encoded in the metric tensor can also
be re-expressed in terms of the {\em distance} function
\begin{equation}\label{distance}
d(x,y) := \inf_{\sigma(0)=x,\ \sigma(1)=y} \left( \int_0^1 |\dot \sigma_t |_{E_{g}}^q dt \right)^{1/q} \qquad \re q > 1;
\end{equation}
the infimum is over smooth curves $t \in [0,1] \mapsto \sigma_t \in M$,  and its value turns out to be independent of $q$ in the range $q>1$.
If instead the metric tensor has a single positive eigenvalue at each point --- schematically denoted by $g\sim (+1, -1, \ldots, -1)$ ---  
the manifold is called {\em Lorentzian} and its metric defines a hyperbolic norm on each tangent space $T_xM$.
Assuming the manifold topology is Hausdorff and the future cone $F_g$ can be chosen to vary continuously throughout $M$,  the manifold is called a smooth {\em spacetime}
when $k=\infty$ (and a $C^k$-smooth spacetime otherwise). 
Under conditions milder than the global hyperbolicity recalled below, 
its {asymmetric} geometry can alternately be encoded in the {\em time-separation} function
\begin{equation}\label{time-separation}
{\gr \ell(x,y)} := {\gr \sup_{\sigma(0)=x,\ \sigma(1)=y}} \left( \int_0^1 |\dot \sigma_t |_{F_g}^q dt \right)^{1/q} \qquad  \gr 0 \ne q < 1,
\end{equation}
where the supremum is taken over smooth causal (i.e. future-directed) curves, and its value is independent of $q$ in the range $0 \ne q<1$;
we define $\ell(x,y)=-\infty$ unless a causal curve links $x$ to $y$.  In either case \eqref{distance}--\eqref{time-separation},  
extremizers are independent of $q$; they are called {\em geodesics}.  The time-separation satisfies a backwards triangle inequality
\begin{equation}\label{rti}
 \ell(x,z) \ge \ell(x,y) + \ell(y,z)
 \end{equation}
for all $x,y,z \in M$, analogous to the usual {triangle inequality} satisfied by a Riemannian distance $d$.

{\em The Riemann curvature tensor:}
given timelike (future-directed) {\em geodesics} $\gr (\sigma_s)_{s\in[0,1]}$ and $\re (\tau_t)_{t\in[0,1]}$ with ${\gr \sigma_0}={\re \tau_0}$ and
 $\dot \tau_0 - \dot \sigma_0 \in F\setminus \p F$ in a $C^2$-smooth spacetime,  Taylor expanding the time-separation function on $t>s$ yields
$$
\ell({\gr \sigma_{s},\re \tau_{t}})^2 = | {\re t \dot \tau_0}-{\gr s \dot \sigma_0}|_{F_g}^2 - \frac{\Sec}{6} {\gr s^2}{\re  t^2} + o(( s^2+ t^2)^2)\quad {\rm as}\quad s^2 +t^2 \to 0,
$$
where the  sectional curvature $\Sec=R({\gr \dot \sigma_0, \re \dot \tau_0, \gr \dot \sigma_0, \re \dot \tau_0})$ is quadratic in ${\gr \dot \sigma_0} \wedge \re \dot\tau_0$
and measures the leading order correction to Pythagoras' theorem; the error improves to  $O((s^2 + t^2)^{5/2})$ if the spacetime is $C^3$-smooth.
Polarization of the quadratic form $\Sec$ defines the {\em \re Riemann} tensor 
$R(\cdot,\cdot,\cdot,\cdot)$,  whose trace $\re \Rc_{ik} = g^{jl} R_{ijkl}$ yields the {\em \re Ricci} tensor
$\Rc(v,v)$, which in turn measures the correction to Pythagoras' theorem averaged over all triangles including side $v$.  A second trace $R= g^{ik} \Rc_{ik}$ of the Riemann tensor
yields the {\em scalar curvature} that also appears in Einstein's field equation \eqref{EFE}; on a Riemannian manifold, $R$ gives the leading order correction to the area of a sphere of radius $r$ (and to the volume of a ball of radius $r$) relative to the Euclidean case.  We shall also have need for the Lorentzian volume, which takes the form $d\vg(x) = \sqrt {|\det (g)|} d^n x$ in coordinates;  in the Riemannian case, the same formula gives the $n$-dimensional Hausdorff measure associated to $d$.

\subsection*{Energy conditions, causality, and singularity theorems}

Having now introduced the spacetime geometry and its curvature tensors,  we recall the {\em energy} conditions that play a role 
in the singularity theorems of Hawking and Penrose:

{WEC} ({\em weak energy condition}):  $\gr T(v,v) \ge 0$ for all {\re future $v \in F$};

{SEC} ({\em strong energy condition}): $\re \Rc(v,v) \re \ge 0$ for all {\re future $v \in F$};

{NEC} ({\em null energy condition}): \qquad  $''$  \qquad $\ge 0$ for all {\gr lightlike $v \in \p F$}.
\\ When the cosmological constant $K$ is nonvanishing,  as for dark matter,  the vanishing right-hand side would be replaced by $\ge (n-1)K g(v,v)$.
Neither the strong energy condition nor the weak energy condition implies each other, but either implies the null energy condition, which is expected
to be satisfied by all classical (i.e.~non-quantum) forms of matter \cite{Carroll04}.   
The {\em dominant energy condition} (DEC) --- which we shall not discuss further ---
implies (WEC) and has been interpreted to mean that information cannot propagate faster than the speed of light \cite{HawkingEllis73}.

An {\em inextendible} curve refers to a smooth causal curve $\sigma:(a,b) \longrightarrow M$ defined on an interval $(a,b) \subset \R$ such that neither
$$
\lim_{t \downarrow a} \sigma_t
\quad {\rm nor} \quad
\lim_{t \uparrow b} \sigma_t
 $$
 exists in $M$. A {\em Cauchy surface} refers to a subset $\Sigma\subset M$ meeting each inextendible curve precisely once.
 A spacetime is said to be {\em globally hyperbolic} if a Cauchy surface exists.
Hawking's singularity theorem can be summarized as follows~\cite{Hawking66a}: if a spacetime satisfying
the strong energy condition admits a Cauchy surface with uniformly positive future-directed mean curvature $H_\Sigma \ge h >0$,
then
$$
\sup_{(x, y) \in M \times \Sigma} \ell(x,y) \le 3/h < \infty.
$$
In other words,  an instantaneous lower bound for the rate of expansion of the universe on $\Sigma$ provides a global 
upper bound on the age of any curve until it passes through $\Sigma$,  
hence provides an open class of geometries in which big-bang type singularities are inevitable.  An analogous theorem was proven 
by Cavalletti and Mondino \cite{CavallettiMondino24} in (nonsmooth) globally hyperbolic Lorentzian length spaces \cite{KunzingerSaemann18} 
satisfying the timelike curvature-dimension condition $\TCD(0,N)$ which they introduced in analogy with $\CD(0,N)$.

Penrose' singularity theorem can be summarized as follows \cite{Penrose65a}: if a smooth spacetime with a noncompact Cauchy surface satisfies the null
energy condition, and admits a compact codimension $2$ surface $S$ whose lightlike mean-curvatures are all positive,
then no null geodesic passing through $S$ can be affinely parameterized over the whole real line; it provides an open class of geometries possessing
incomplete geodesics,  like those seen in the spherically (and axi)symmetric black hole solutions of Schwarzschild \cite{Schwarzschild03} (and Kerr \cite{Kerr63}).
While no Penrose type theorem is yet known \cite{Ketterer24} \cite{McCann24} in a Lorentzian length space setting,  Graf \cite{Graf20}
has established a version which holds on any $C^1$-smooth spacetime. 

\subsection*{Smooth Lorentzian splitting theorems}

Let us now recall a Lorentzian analog of the Cheeger--Gromoll splitting theorem,  as conjectured by Yau \cite{Yau82} in the year he 
received his Fields medal,
and proved eight years later by Newman \cite{Newman90}, building on work of others.  
In this theorem,  a {\em line} refers to a doubly-infinite, maximizing, timelike unit-speed geodesic.  A smooth spacetime is called  
{\em timelike geodesically complete} if all unit-speed timelike geodesics admit doubly-infinite extensions  which are locally maximizing everywhere but not necessarily globally maximizing.

\begin{theorem}[Lorentzian splitting \cite{Newman90} conjectured in \cite{Yau82}]
\label{T:Newman} 
Let $(M^n,g_{ij})$ be a connected smooth spacetime satisfying the strong energy condition (SEC) and containing a timelike line.
If $M$  is {\gr (a) timelike geodesically complete,} 
then $M$ is a geometric product of $\R$ with a (Ricci nonnegative, complete) Riemannian submanifold $\Sigma^{n-1}$.
\end{theorem}

The same conclusion had already been deduced assuming $(M^n,g_{ij})$ admits a compact Cauchy surface by Galloway \cite{Galloway84},
under sectional curvature bounds assuming {(b) global hyperbolicity} by Beem, Ehlich, Markvorsen and Galloway \cite{BeemEhrlichMarkvorsenGalloway85},
and then under timelike Ricci nonnegativity by Eschenburg \cite{Eschenburg88} assuming (a)--(b)  and finally by Galloway \cite{Galloway89} assuming (b) global hyperbolicity without (a).

Like the Cheeger--Gromoll proof of the Riemannian splitting, most of these works employ a Lorentzian analog of the Busemann function which can be defined as follows.
Letting $\gamma:\R\longrightarrow M^n$ be the isometrically embedded proper-time parameterized line, set
\begin{align*}
b^+_r(x) &:=-\ell(x,\gamma(r)) +\ell(\gamma(0),\gamma(r)) 
\\ b^-_r(x) &:=\phantom{-}\ell(\gamma(r),x) - \ell(\gamma(r),\gamma(0)) 
\end{align*} 
and 
$$b^\pm :=\displaystyle \lim_{r \to \pm \infty} b^\pm_r.
$$ 
Then $b^\pm_r$ is $1$-steep --- meaning $b^\pm_r(y) - b^\pm_r(y) \ge \ell(y,x)$ for all $x,y \in M$ --- and $|\nabla b_r|_F=1=|\nabla b^\pm|_F$ whenever these derivatives exist.
For $r>0$, the reverse triangle inequality \eqref{rti} again gives 
\begin{equation}\label{ordering}
\gr b^+_r \ge b^+ \ge b^- \ge -b^-_{-r}
\end{equation}
for $r>0$ with equality at $x=\gamma(0)$.  At this point however,  the classical Lorentzian proofs are forced to diverge from the strategy of Cheeger and Gromoll
since --- unlike Calabi's theorem \cite{Calabi58v} --- the d'Alembert comparison theorem of Eschenburg \cite{Eschenburg88} was not known to extend across
the timelike cutlocus nor to survive the limit $r\to \infty$;  moreover,  without a maximum principle one cannot conclude from equality at $x=\gamma(0)$ in the ordering 
\eqref{ordering} that the
$p$-super- and $p$-subharmonic functions $b^+ \ge b^-$ coincide;  finally in the spacetime setting the leftmost expression in Bochner's identity \eqref{Bochner} is no longer nonnegative definite.  All three failures stem from the lack of ellipticity of the Lorentzian Laplacian $\square_2$ --- better known as the {\em d'Alembertian} or {\em wave operator}.
Thus previous researchers have been forced to perform crucial steps of their analyses on well-chosen spacelike hypersurfaces --- such as a level set $\{b^+=0\}$ as in \cite{Eschenburg88} or a zero mean curvature submanifold provided by Bartnik~\cite{Bartnik88} as in \cite{Galloway89} --- on which ellipticity is restored, before propagating the information so gleaned backwards and forwards in time.

The purpose of this lecture is to describe a new approach for proving the Lorentzian splitting theorems, 
developed in joint work with the quintet \cite{BraunGigliMcCannOhanyanSaemann24+}. 
In this approach we sacrifice linearity of the d'Alembertian to gain ellipticity,  which allows us to hew more closely to the 
Riemannian strategy of Cheeger and Gromoll \cite{CheegerGromoll71}.  For smooth spacetime metrics $g_{ij} \in C^\infty(M^n)$,
precise statements can be found in \cite{BraunGigliMcCannOhanyanSaemann24+},  though our conclusions will be extended to less smooth metrics
$g_{ij} \not\in C^2(M^n)$ in a subsequent work.  A central role in our analysis is played by the $p$-d'Alembert operator 
$\square_p u := - \nabla \cdot (|\nabla u|_{F}^{p-2} \nabla u) = -\frac{\delta E}{\delta u}$ 
in the negative-homogeneity range $p<1$.
Here $u$ is assumed to be future-directed --- meaning $u(y)\ge u(x)$ if $\ell(x,y) \ge 0$ ---  and the operator arises as the variational derivative
of the energy 
$$
E(u) = \int_M H(du) d\vg
$$
induced by the Hamiltonian $H(w) = -\frac1p |w|_{F^*}^{p}$.  Nonuniform ellipticity of $\square_p u$ follows from the convexity of 
this Hamiltonian established for $p<1$ by McCann~\cite{McCann20} and alternately by Mondino and Suhr~\cite{MondinoSuhr23}.  The convex Lagrangian
$L(v)=-\frac1q |v|_F^q$ satisfies $DH=(DL)^{-1}$ if $p^{-1} + q^{-1}=1$.  Even on smooth Lorentzian spacetimes,
the Lagrangian (and Hamiltonian) are defined to be $+\infty$ outside the future cone $F \subset TM$ (or its convex dual cone $F^*\subset T^*M$ respectively); 
in the subrange $p<0$ --- or equivalently $\re 0<q < 1$ --- the Lagrangian
$L$ jumps from $0$ to $+\infty$ across the
boundary  $\p F$ , while the Hamiltonian
$H$ diverges continuously at the boundary $\p F^*$ of the dual cone.

A comparison theorem for this operator was recently established in the (b) globally hyperbolic (but nonsmooth) timelike curvature-dimension
setting $\TCD(0,N)$ of \cite{CavallettiMondino24} jointly with the octet \cite{BeranOctet24+}:

\begin{theorem}[Nonsmooth $p$-d'Alembert comparison \cite{BeranOctet24+} \cite{BraunGigliMcCannOhanyanSaemann24+}]\label{T:p-d'Alembert comparison}
For $p<1$, the operator $\square_p u := - \nabla \cdot (|\nabla u|_{F}^{p-2} \nabla u)$ is {\re nonuniformly elliptic}
on the set of future-directed functions $u$,
and (SEC) implies $\re \square_p b^+_r \le \frac{n-1}{\ell (\cdot, \gamma(r))}$ distributionally, meaning for all $0 \le \phi \in C^1(M)$ with compact support,
\begin{equation}\label{p-d'Alembert comparison}
\int_M  g\left({\nabla  \phi},\frac{\nabla b^+_{\re r}}{|\nabla b^+_{r}|_F^{2-p}} \right) d\vg \le
 (n-1)\int_M \frac{{\phi(\cdot)} d\vg(\cdot)}{\ell(\cdot,{\gamma(r)})}. 
\end{equation}
\end{theorem}

A logically independent and much simpler proof of \eqref{p-d'Alembert comparison} in the (a) timelike geodesically complete smooth case has been given by the quintet \cite{BraunGigliMcCannOhanyanSaemann24+}. 
An important ingredient in the latter proof is Eschenburg's $2$-d'Alembert comparison inequality --- which holds outside the timelike cutlocus and can also be recovered from \eqref{p-d'Alembert comparison} since the approximate Busemann functions satisfy $|\nabla b^\pm_r|=1$ a.e.  Moreover, to obtain equality  $b^+=b^-$ 
of the super- and subsolutions from their ordering \eqref{ordering} and tangency along $\gamma$,  we need to improve the first conclusion of Theorem~\ref{T:p-d'Alembert comparison}
by establishing uniformity for the ellipticity of $\square_p b$ at $b=b^\pm$. To get this uniform ellipticity near $x \in M$ requires bounding {\gr $\{\nabla b^+_r(x)\}_{r \ge R}$ away from the lightcone asymptotic to the noncompact pseudosphere $|w|_F=1$. Indeed, the linearization 
\begin{align*}
\square_p b &= \nabla_i (\frac{\p H}{\p w_i}\bigg|_{db})
= {\re H^{ij}} \nabla_i \nabla_j b 
\end{align*}
of the operator in non-divergence form involves the Hessian
\begin{align*}
 {\re H^{ij}} := \frac{\p^2 H}{\p w_i \p w_j} &= 
|w|^{p-2}\left[ {\gr (2-p) g^{ik} g^{jl} \frac{w_k w_l}{|w|^2}} {\re-  g^{ij}}\right]
\\ & \sim  |w|^{p-2} 
\left[\begin{array}{cccc} {\gr 2-p} {\re -1} &0&0&0 \\ 0 & {\re 1} &0&0 \\  0&0& \cdots &0 \\ 0&0&0& {\re 1} \end{array} 
\right],
\end{align*}
which becomes positive definite if $p<1$ provided we can choose
{normal coordinates} around $\re \gamma(0)$ in which $w=db$ is the time axis.

Although the linearization above is heuristic, uniform ellipticity can be rigorously established in divergence (rather than non-divergence) form 
using a result first proved by Eschenburg 
assuming (a)--(b), and for which a simpler proof was found later by Galloway and Horta 
assuming either (a) or (b).  To formulate it, recall that any smooth spacetime $(M,g)$ admits a complete Riemannian metric tensor $\tilde g$ according
to Nomizu and Ozeki \cite{NomizuOzeki61}. 

\begin{theorem}[Equi-Lipschitz estimate 
\cite{Eschenburg88} \cite{GallowayHorta96}]
Under (a) and/or (b),
 $\gr \gamma(0)$ admits a neighbourhood $\gr X$ and constants $\re R, C$ such 
 that if $\re r \ge R$ then (i) a maximizing geodesic $\gr \sigma$ connects each $\gr x \in X$ to $\re \gamma(r)$;
 (ii) each such geodesic satisfies 
 $\gr \tilde g(\sigma'(0),\sigma'(0)) \le {\re C} g(\sigma'(0),\sigma'(0))$ 
 hence
 $\{b^+_r\}$ is timelike and uniformly equi-Lipschitz on $\gr X$. 
\end{theorem}

Intersecting the ellipsoid $\tilde g(w,w) \le C$ with the hyperboloid $g(w,w) \ge 1$ prevents $db$ from approaching the light cone,
hence uniformizing the ellipticity of $\square_p b^+$ on $X$. However, to deduce \eqref{p-d'Alembert comparison} when $r=\infty$ it is not enough
that $b_r^+ \to b^+_\infty$ locally uniformly;  we shall also need $\nabla b^+_{r} \longrightarrow \nabla b^+$ a.e.  We get this convergence by 
controlling one more derivative than the previous theorem:

\begin{lemma}[Equi-semiconcavity 
\cite{BraunGigliMcCannOhanyanSaemann24+}]
 For some constant $\tilde C$, all $u \in \{b^+_r\}_{r \ge \re R}$ and $(v,x) \in T \gr X$ satisfy
\begin{align*}
\re \lim_{t \to 0} \frac{u(\exp_x^{\tilde g} t v) + u(\exp_x^{\tilde g} -tv) -2 u(x)}{\tilde g(v,v)} \le \tilde C
\end{align*}
\end{lemma}

Equipped with this lemma,  the $p$-d'Alembert comparison result \eqref{p-d'Alembert comparison} established by Eschenburg
\cite{Eschenburg88} where $b^\pm_r$ is smooth can be extended across the timelike cutlocus and to $r=\infty$.
Thus $\pm b^\pm$ are distributionally $p$-superharmonic $\re \square_p b^+ \le 0 \le \square_p b^-$;
moreover, the strong maximum principle now improves $b^+\ge b^-$ to $\re b^+ =b^- \in C^{1,1}({\gr X})$ \cite{BraunGigliMcCannOhanyanSaemann24+}.
A homogeneity $2p-2<0$  variant on Bochner's identity \eqref{Bochner} 
derived by the quintet (and which can alternately be viewed as a special case of Ohta \cite{Ohta14h}, 
or see the appendix of Mondino and Suhr \cite{MondinoSuhr23}) reads
\begin{align*}
 & \Tr\left[ \left( \sqrt{D^2 H} \nabla^2  b \sqrt{D^2 H} \right)^2\right] + \re \Rc(DH,DH)
\\ & =  H^{ij} b_{jk} H^{kl} b_{li} + R_{ij} H^i H^j
\\ & =   \nabla_i (H^{ij}|_{d b} \nabla_j ({\re H|_{d b}})) - H^i \nabla_i ({\re \nabla_j (H^j |_{d b})})
\\ &= 0,
 \end{align*}
 where the final equality follows from the identities $|db|_{F^*}=1$ and $\square_p b=0$ satisfied by the Busemann function $b:=b^\pm$
 on $X$.  Unlike the Lorentzian metric $g^{ij}$,  the Hessian matrix $H^{ij}$ of the Hamiltonian is positive-definite, therefore allowing us to deduce
 $\Hess b =0$ in $X$ from the timelike Ricci nonnegativity hypothesis (SEC).
 
 As in the Riemannian case, $\Hess b =0$ implies $\nabla b$ is a timelike Killing vector field whose flow gives a local isometry on $X$,
 and moreover that $\Sigma_r := \{ x\in X \mid b(x)=r\}$ is totally geodesic since its normal $\nabla b$ is parallel.  Thus on $X$, the metric 
 $g_{ij}$ splits orthogonally into components tangent $g_{ij} dy^i dy^j<0$ and normal $dr^2$ to $\Sigma_r$.  Because the same argument can be made
 at each point $\gamma(r)$ of the line,  one can extend the neighborhood $X$ of $\gamma(0)$ to a neighbourhood $\tilde X$ of the entire line $\gamma(\R)$.
 
Since the uniform ellipticity is local,  it is not obvious that $b^\pm$ agree or are finite outside $\tilde X$.  Additional arguments are therefore required to 
conclude first that $\tilde X$ can be take to be invariant under the flow of $\nabla b$ (unlike the snake which swallowed an elephant drawn in Le Petit Prince
\cite{Saint-Exupery43}),
and finally that $\tilde X = M$ by connectedness of $M$.  However, these arguments can be patterned on the original proofs 
\cite{Eschenburg88} 
\cite{Galloway89} 
\cite{Newman90}, and are actually simpler in the (a) timelike geodesically complete case of 
Newman than in the (b) globally hyperbolic case of Galloway; see \cite{BraunGigliMcCannOhanyanSaemann24+} for details.

\end{document}